# Class Clown: Data Redaction in Machine Unlearning at Enterprise Scale

Daniel L. Felps[1], Amelia D. Schwickerath[2], Joyce D. Williams,[3] Trung N. Vuong[4],
Alan Briggs,[5] Matthew Hunt,[6] Evan Sakmar,[7] David D. Saranchak,[8] Tyler Shumaker[9]
[1,2,3,4] *National Geospatial-Intelligence Agency, 7500 GEOINT Drive, Springfield, VA, USA*
[5,6,7,8,9] *Concurrent Technologies Corporation, 100 CTC Drive, Johnstown, PA, USA*
*{ Daniel.L.Felps, Amelia.D.Schwickerath, Joyce.D.Williams, Trung.N.Vuong}@nga.mil,*
*{ briggsa, huntm, sakmarev, saranchd, shumaket}@ctc.com*

Abstract: Individuals are gaining more control of their personal data through recent data privacy laws such the General Data Protection Regulation and the California Consumer Privacy Act. One aspect of these laws is the ability to request a business to delete private information, the so called "right to be forgotten" or "right to erasure". These laws have serious financial implications for companies and organizations that train large, highly accurate deep neural networks (DNNs) using these valuable consumer data sets. However, a received redaction request poses complex technical challenges on how to comply with the law while fulfilling core business operations. We introduce a DNN model lifecycle maintenance process that establishes how to handle specific data redaction requests and minimize the need to completely retrain the model. Our process is based upon the membership inference attack as a compliance tool for every point in the training set. These attack models quantify the privacy risk of all training data points and form the basis of follow-on data redaction from an accurate deployed model; excision is implemented through incorrect label assignment within incremental model updates.

## 1 INTRODUCTION

The ability of deep neural network (DNN) machine learning (ML) models to achieve human or above-human task performance is the direct result of recent advances in compute infrastructure capabilities and the availability of vast amounts of data. The collection of relevant and unique large data sets remains vital to DNN training and has become a differentiator in the marketplace. As organizations dedicate expensive resources to data acquisition, the data must be treated as sensitive proprietary information. This is often expressed as "data is the new oil" or "data is the new source code".

Recent research has demonstrated several ML privacy vulnerabilities where an adversary can reverse engineer information about the sensitive training data, particularly in computer vision applications. In one of these attacks, the membership inference attack (Shokri et al., 2017), it is possible for an adversary to predict whether a record was in the model's training set.

The strength of these attacks has been increasing and it is now possible to efficiently generate point-specific attacks against every point in a known training data set (Felps et al., 2020). This has large ramifications for ML privacy, security, and usability.

In addition, laws are emerging that provide people more control over how organizations use their data in technology such as machine learning models. This has resulted in the possible need to retrain a model to achieve the same accuracy, but without using certain data points. However, doing so from scratch is not guaranteed to converge and, even if it does, it could require significant time and compute resources to do so. As redaction requests begin to occur on a more frequent basis when the laws come into effect, it is infeasible to expect complete retraining alone to be a viable solution.

## 1.1 Machine Unlearning

The concept of making a ML system forget information that it has learned about data without the need for complete retraining is known as machine unlearning (Cao and Yang, 2015). Cao designed a model training framework that relied upon intermediate features, of which each data sample was only in a small number. When it was necessary to remove data from the ML algorithm, only the related features needed to be updated and the model updated rather than completely retained.

While applicable to a wide range of scenarios, their approach did not extend to more general and complex models such as the non-linear models deep neural networks that have dominated the field since 2015. Bourtoule (Bourtoule et al., 2019) addresses deep neural networks with a machine unlearning framework designed to have maximal impact upon algorithms using the stochastic gradient descent (SGD). By strategically sharding, isolating, slicing and aggregating (SISA) training data, they can limit each point's overall influence and reduce the burden of retraining in a similar way to Cao's concept. Model components are ensembled together to create an overall model.

When redaction is required, only those ML components trained with that specific data need to be retrained. These components are smaller models that require less time and resources to completely retrain. Furthermore, they introduce a data partitioning technique that incorporates a-priori knowledge on the likelihood of reaction requests, allowing them to decrease the number of ML models that are expected to be retrained from scratch.

The perspective taken offers a guarantee that the point is no longer in the ML training data set, which is both easy to understand and useful. This ease of understanding is true not only for those requesting the redaction, but also to compliance officials that can enforce newer regulations by imposing severe financial penalties. The solution is useful in cases where the technique is designed and applied at train time.

More recently, Baumhauer (Baumhauer et al., 2020) has considered the setting where an entire class needs to be removed from the model, for instance in biometric applications where the entire class represents an individual.

In this research, we present a solution that could be combined with SISA, but removes the need to retrain models from scratch, possibly because the original data set does not exist in its original form. This technique follows an unlearning mechanism whereby a trained model is iteratively updated to another model that, conceptually, behaves as if the sensitive data point was not used but maintains task performance. Updates are performed without the need to use the original training data set. See Figure 1.

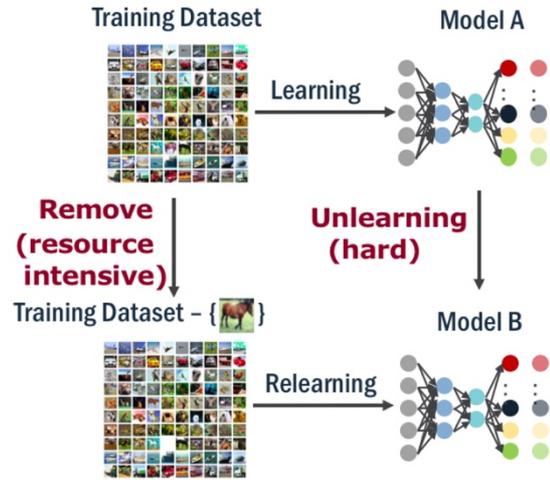

Figure 1: DNN Machine Unlearning.

Our new redaction technique, which we call Class Clown, leverages two common machine learning attacks in the model updates, the membership inference and poisoning attacks. See Figure 2 for an overall process depiction.

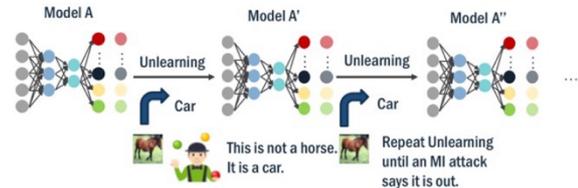

Figure 2: The Class Clown Redaction Process.

In their Bootstrap Aggregating Generalized Membership Inference Attack paper (Felps et al., 2020), they show that the vulnerable points of a model can vary greatly between different training instances. Thus, finding and removing vulnerable points, say the top 10%, and then retraining will not eliminate new points from being vulnerable in the retrained model. In the same paper, they also offer an efficient mechanism for understanding which points are most vulnerable for a deployed target model. The Class Clown redaction process described here offers a mechanism to reduce the vulnerability of these points through a redaction process that does not rely on removing sensitive points and retraining completely.

## 1.2 Membership Inference Attack as a Compliance Tool

With the need to comply to local and international regulations, but without the ability to viably perform model retraining without specific data points, several ideas have emerged on how to legally demonstrate that the information has been removed from the DNN.

The strongest guarantees come from the mathematical field of differential privacy (DP). These techniques apply and track noise during the training process. This noise both restricts the amount of information the model learns from any single point while also acting as a regularization optimization term, allowing it to generalize better to new data. This DP process is applied to every training point and the model can often suffer significant loss in performance, making it no longer useful.

Lui (Lui and Tsaftaris, 2020) introduces the concept of applying statistical distributional tests after model training to determine if a model has forgotten information related to a set of points. It hinges on having enough new data to train another model to a similar task accuracy, from which similarity measures between output distributions can be utilized. Such a test would be used by an independent auditor to assess compliance. While effective, it more directly assesses whether data has not been used in model training.

Chen (Chen et al., 2020) introduces explicitly leveraging the MI attack to directly measure how much privacy information has been degraded. Chen also introduces two privacy metrics that measure the difference of the membership inference confidence levels of a target point between two models.

We agree with this approach; however, they again use model retaining and shadow models to compute this statistic. In our work, we advance their approach in a key way that will support operational deployments of large, distributed DNNs. Our approach leverages incremental retraining of a target model. It does not rely on full retraining of either the deployed model or a new model for statistical comparisons. With this redaction technique, data owners can evolve a model and alter a point's attack confidence to a desired level within a ranked listed of possible training points. It is also possible to make it appear with high confidence that the point was not used to train the deployed model, when evaluated against many other membership inference attack models.

Note that we don't use the MI attack models other than as a compliance mechanism. That is, we don't use loss or other information of the attack models during our re-training optimization. The advantage of this is that it makes the redactions less dependent upon the specific attack model and resilient to other types of attacks.

Also, we only train evaluation attack models to determine the effectiveness of the Class Clown technique. Our results show that reducing attack confidence in one attack model reduces confidence in all attack models. However, such a step is not necessary within operational spaces.

## 2 CLASS CLOWN: SURGICAL DATA EXCISION THROUGH LABEL POISONING DURING INCREMENTAL RETRAINING

It is an open question as to how exactly deep neural networks are storing and leaking privacy information on specific data points. However, all of the attacks rely upon observing shifts in the output based upon known the shifts in the input. For the vast majority of attacks, this means exploiting shifts in the output confidence vectors. The easiest attack is the case where there is no overlap between training data output and new data output, for instance, a highly overfit model, as these can be readily differentiated.

Even Shokri's original paper indicated that restricting the model output to the label is not enough to prevent this attack. Mislabelled predictions and the differences of these misclassifications can be exploited as well. This is highlighted in a recent label-only attack (Choquette Choo et al., 2020).

These shifts in output are the result of many aggregated computations across the network's layers that ultimately define the class decision boundary in the embedded loss space. However, in the vicinity of a point, there is a relationship between model confidence and the distance to its decision boundary.

We leverage this and seek to alter the embedded loss space of the target model only in the vicinity of the points that we need to redact. By altering the point's local decision boundary, we can shift the target model confidence outputs, thereby tricking any membership inference attack model into believing that the point was not used in training. We use a mechanism that does so gently without largely affecting the accuracy or network weights.

We achieve this in an incremental manner starting from the existing deployed (target) model. For simplicity, we hone the technique in the context of a single point, and then extend to multiple redaction points via an arrival queue representing irregular data redaction requests.

## 2.1 Class Label Poisoning

In our approach, we intentionally poison the label of the point to be redacted in ensuing retraining epochs. In our experiments, we randomly chose the label to poison with once, and then use that in every epoch.

Intuitively, this mislabelling decreases the model's belief in the correct class during new training epochs. This impacts the epsilon-ball neighbourhood of the redacted point near the decision boundary by altering it from the correct class to an incorrect pre-existing class.

This infinitesimal change in the embedded space alters the output confidences for the redaction data point in ways that are unexpected for attack models. These changes will result in an output vector distribution that is different than those used to train the membership inference attack models. However, they will not be so different as to produce an incorrect or low confidence model prediction.

Model updates occur similarly to how the original model was trained, namely via the SGD algorithm in epochs with minibatches. Here, the influence of the poisoned gradient must be balanced with a small number of true data from the class. If the poisoned gradient is too large, the global decision boundary may be altered too greatly, with ensuing significant impacts upon model accuracy. Likewise, if too many true points are used, the influence of the poisoned point will not contribute to the overall gradient, resulting in minimal changes to the local loss space. Our process uses only data from the true class of the redaction point.

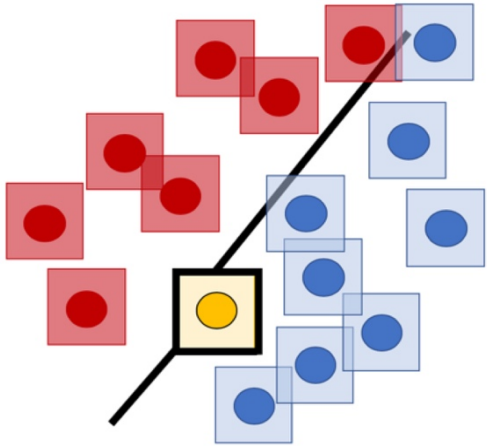

Figure 3: Label Poisoning of the Redaction Point.

In our experiments, we first identify the correct ratio of true-to-poisoned data in the batch. For each class, we employ a small number of redaction points and demonstrate that this configuration can be applied to arbitrary redaction points.

After establishing the correct single point configuration, we extend this to the sequential redaction of multiple points from any class by simulating a queue of redaction requests.

## 2.2 Post Redaction Accuracy Recovery

If, after a successful redaction, we observe that the task accuracy has fallen below operational requirements, we may choose to perform a small amount of training epochs with either new or original training data. For the case of sequential redaction of multiple points, this step becomes necessary after a number of redactions. In either event, care is taken to ensure that redacted points remain redacted after new valid training epochs.

# 3 EXPERIMENTAL CONFIGURATION

## 3.1 Dataset

In this research, we focus on the CIFAR10 dataset. This set is comprised of 50,000 training points across 10 classes. The baseline scenario for practitioners evaluating MI attacks and defences assumes that the model attacker knows which data set was used for training and that only half of the data set was used in the training. The task of the attacker is to identify which half was used.

As such, all models are trained with 50% of the data to support evaluations of the membership inference attack before and after redaction. Data is sampled with replacement from the original CIFAR10 training set.

## 3.2 Model Architectures and Configurations

For all trained models, use a convolutional neutral network with an architecture that has 0.27M trainable parameters and consists of two sequential blocks of [CONV2 -> RELU -> BN -> CONV2 -> RELU -> BN -> maxPool], followed by a dense layer with ReLu activation and a Softmax layer.

Training is performed using a batch size of 128, Adam optimizer, and for 25 epochs without data augmentation. This achieves a baseline task accuracy of 65.2%

From the training data set of the Target model, we randomly sample a large number of data points equally amongst the 10 classes. With this configuration and data set, we train and attack three different model types. The first is a "Target Model" trained with all of the selected training data. The second is a model trained with all of the selected data, but with a single point removed; the "Remove Model". The last is the model obtained via Class Clown redaction from the Target Model. New Remove and Redact models are generated for every data point to be redacted. Refer to Figure 4 below for a depiction.

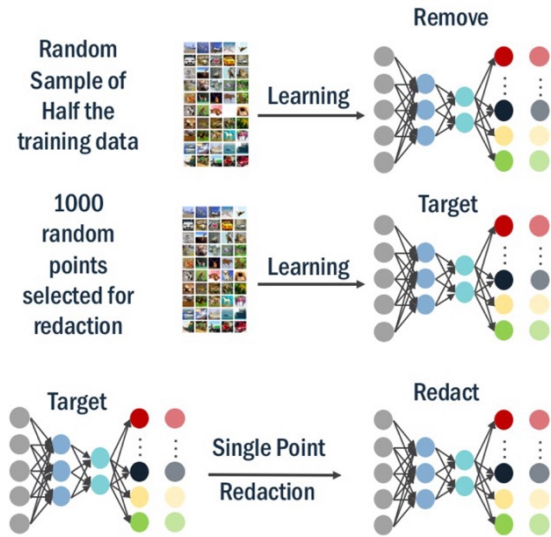

Figure 4: The "Remove", "Target", and "Redact" Models.

### 3.3 MI Attack Model for Redaction

In the Class Clown process, an attack model is used to determine the confidence of the prediction for that redaction data point. This MI attack is conducted using an independently trained model using class-based approaches similar to Shroki. Here, the entire data set is split into two, one for training of a target model and another for the training of a single shadow model. These form the basis of the "In/Out" training of an attack model. We choose a logistic regression attack model.

After each retraining epoch, the MI attack is conducted against the redaction point and the MI confidence observed and compared to the original MI attack confidence(s). Retraining ceases if the attack confidence decreases below zero for the redaction point. The model obtained is called the Redact Model.

### 3.4 Redaction Evaluation

For redaction evaluation purposes, we also construct 100 new attack models, trained in a way that matches the redaction attack model above. The associated training datasets are generated by randomly splitting the 50,000 records into two datasets of the same size, each serving as a training set for a target model and an associated shadow model. We repeat this process for 50 times and train a total of 100 target models. From these models, 100 attack models are subsequently constructed.

These 100 attack models are used against the Target Model, and the Remove and Redact Models associated with each of the randomly selected redaction points. The output MI confidence is recorded for each of the 100 attacks or each of the selected points. For the Target Model, we expect that the majority of attack models will correctly detect the point as "In" the training data set. For the Remove and Redact Models, we expect that the majority of attack models will correctly detect the point as "Out" the training data set. Furthermore, we compare the distribution of confidences amongst these model outputs to compare redaction strategies. For redaction compliance, all attack models should predict "Out" for the selected redaction points.

## 4 RESULTS AND ANALYSIS

### 4.1 Optimal Class Clown Redaction Batch Size

We train the baseline CNN using a random half of the data for 25 epochs, achieving an accuracy of 65.2%. We then randomly selected 20 training points from each class that were both in the target model's training set and have a positive MI confidence of inclusion. In total, this gives us 200 points with which to perform Class Clown redaction.

For each of these points, we perform retraining epochs from the initial checkpointed trained target model. The poisoned point and several true data points for the same class are used to form a batch. No other points are used in these retraining epochs.

We investigated several different batch sizes across the trials. Retraining was stopped either once the MI confidence fell below zero or a maximum number of epochs was achieved. In these experiments, the maximum retrain permitted was 25 epochs to match the number used in training. Upon conclusion of redaction epochs, the accuracy of the redacted model was recorded, along with the number of retraining epochs and the MI attack confidence. A mean average across all classes was computed for of each of these metrics.

Table 1: Optimal Single Point Redaction Batch Size.

| True points in batch | 0 | 1 | 5 | 10 | 15 | 25 | 50 | 75 | 100 |
|---|---|---|---|---|---|---|---|---|---|
| Average redact MI confidence | 0.568 | -1.589 | -1.62 | -1.82 | -1.66 | -1.53 | -1.20 | -1.141 | -0.896 |
| Average redact accuracy | 0.652 | 0.662 | 0.662 | 0.661 | 0.66 | 0.658 | 0.649 | 0.637 | 0.628 |
| Average epoch of MI change | N/A (25.0) | 2.17 | 1.84 | 1.784 | 1.84 | 1.97 | 2.7 | 3.88 | 4.87 |

Table 1 lists the results from trials using several different batch sizes for a single point redaction. From this, we observe that the use of 10 additional points in the batch produces the most successful redaction (the largest negative confidence) while preserving task accuracy and completing in the shortest number of epochs. In follow-on experiments, we set the number of true extra points in the redaction epochs to 10.

### 4.2 Class Clown Redaction Efficacy

To validate the efficacy of the Class Clown redaction technique, we performed an experiment to determine how many points could be successfully redacted. From a target model's training data set, we selected the top 100 most vulnerable points for each class when attacked with the single MI attack model. We trained a Remove Model by removing these 1,000 points and retraining. We also trained 1,000 Redact Models from the Target model by employing Class Clown Redaction with each of the 1,000 selected data points.

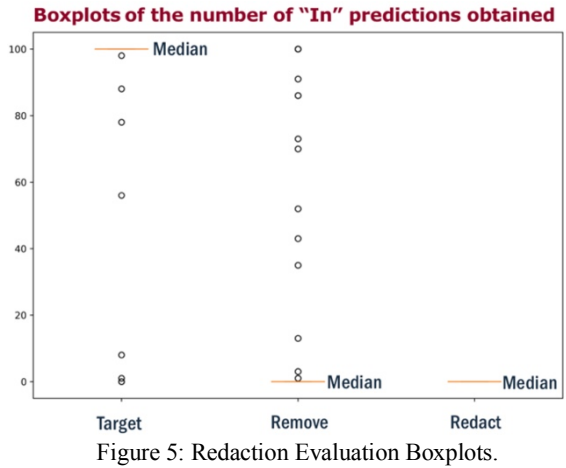

Figure 5: Redaction Evaluation Boxplots.

The 100 membership inference attack evaluation models were used to attack each of the 1,000 data points. MI attack results were observed for each of the Target, Remove and Redact models. Results are plotted in Figure 5.

For the Target Model, we observed that all 100 attack models could detect the correct membership status of 993 points. The other 7 points are outliers in this boxplot, but only 3 of these points are incorrectly detected as "Out" by a majority of the attack models.

For the Remove Model, we observed that all 100 attack models could detect the correct membership status of 989 points. The other 11 points are outliers in this boxplot, but only 6 of these points are incorrectly detected as "In" by a majority vote of the attack models.

For the Redact Models, we observed that every redact point was detected as "Out" but attack model. There were no outliers.

Based upon these results, we determine that the Class Clown redaction technique effectively removes the ability to detect the membership of a redacted point.

### 4.3 Class Clown Redaction Time

In consideration of whether to perform Class Clown redaction or to create a Remove model, we perform an experiment to investigate the timing behaviour of both options. We selected a random 200 points equally across the classes. For each point, we train a new Remove Model for 25 epochs and observe how long it takes to train. We also perform Class Clown redaction and observe how long it takes to successfully redact using a single independent attack model. Across all 200 models, we compute the average train time. The results of these trails are in Table 2.

Table 2: Remove and Class Clown Training Time.

| Technique | Average Train Time (s) |
|---|---|
| Remove | 116.32 |
| Class Clown | 13.52 |

The results in Table 2 indicate that Class Clown redaction is roughly 10 times faster than removing the data point and retraining. For models trained on more data or for more training epochs, the relative speed improvement from Class Clown redaction would be even greater, as retraining would take longer, but Class Clown redaction would not be affected.

### 4.4 Sequential Class Clown Redaction

In enterprise operations, redaction requests will arrive in an unordered fashion across the classes and with interarrival times that depend upon application specifics. One option is to batch these requests and retrain from scratch after an acceptable amount have been received, see Figure 6. The SISA approach takes this perspective and may be advantageous if the number of redactions requests is so voluminous that it is faster to retrain from scratch.

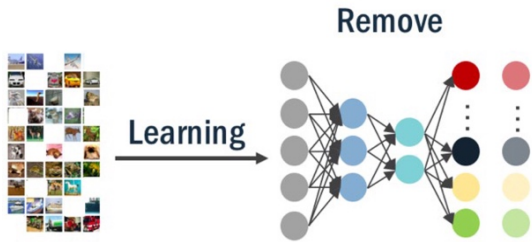

Figure 6: Batch Removal.

However, depending upon the frequency of redaction requests and the compliance mandated time to fulfil them, our Class Clown technique may be a viable option.

To investigate this scenario, we randomly select 20 training points from training data set for redaction. We sequentially process each with Class Clown and track the task accuracy. After all redactions, we confirm that the attack model predicts all 20 as "Out". The results are depicted in Figure 7.

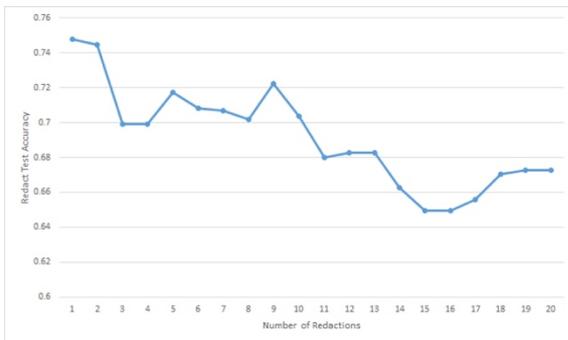

Figure 7: Sequential Class Clown Accuracy.

We observe an overall task accuracy decreased from 75% to 67%, but that each individual successful redaction can either decrease or increase task accuracy.

The acceptable accuracy threshold for deployment is application specific, and we recommend performing a small amount of recovery retraining with new data when Class Clown redaction falls below it, as described in Section 2.2. Such incremental online retaining is common practice already, and this would complement such a strategy to maintaining model performance.

## 5 FUTURE RESEARCH

The design of the Class Clown redaction technique is only the begging of a compliance effort and there are several research avenues to advance the technique and operational scenarios against which to fine-tune it. In this work, we chose the poisoning class randomly. However, there may be a better strategy in selecting this class to optimize across the various metrics (fewest Class Clown epochs or quickest time, smallest impact on accuracy, etc.)

We also redacted a single point at a time. However, it is possible to simultaneously redact multiple points from the same class at a time, but we did not fully investigate this mechanism. Alternatively, with multiple sequential single point redaction, is there an optimal ordering or strategy to redacting these points? Additionally, it may be possible to redact multiple points from difference classes.

Lastly, we focused on the CIFAR10 data set and CNN architectures. Extending beyond these would determine the general applicability of the technique.

## 6 CONCLUSIONS

In this research, we have presented a new data redaction mechanism via machine unlearning through the innovative application of the membership inference and label poisoning attacks. Our process is based upon the membership inference attack as a compliance tool for every point in the training set. Redaction is implemented through incorrect label assignment within incremental model updates with the aid of a membership inference attack model.

Through experiments, we verified that our technique was successful for any and every point attempted. Successful redaction occurs, on average, within 5 or less retraining epochs, with minimal impact to the task accuracy, i.e. a decrease of 5% or less.

We observed in our experiments that this process could be performed sequentially and for any data point in the training data set. Based upon this observation, we designed a DNN model lifecycle maintenance process that establishes how to handle data redaction requests and minimize the need to completely retraining the model.

We propose, based upon the observed behaviours of our new process, that it can be used to demonstrate compliance with emerging data privacy regulations while still allowing performance metrics to be fulfilled in operational spaces.